\documentstyle[pra,aps,multicol,psfig]{revtex}

\def\addcontentsline#1#2#3{\relax}
\begin{document}
\draft  
\title{Umklapp scattering in transport 
through a 1D wire of finite length. }
\author{V.V. Ponomarenko$^{*}$ }
\address{Department of Theoretical Physics, University of Geneva,
24, quai Ernest-Ansermet, 1211 Geneva 4}
\date{\today}
\maketitle
\begin{abstract}

Suppression of electron current $ \Delta I$ through a 1D channel of length $L$ 
connecting two Fermi liquid reservoirs is studied taking into account the 
Umklapp interaction induced by a periodic potential.  
This interaction opens band gaps at the integer fillings  and Hubbard gaps $2m$ at some rational fillings in the infinite wire: $L \to \infty$. In the perturbative regime where $m \ll v_c/L$ 
($v_c:$ charge velocity), and for small deviations $\delta n$
of the electron density from its commensurate values 
$- \Delta I/V$ can diverge with some exponent as 
voltage or temperature $V,T$ decreases above  $E_c=max(v_c/L,v_c \delta n)$,
while it goes to zero below $E_c$.
This results in a non-monotonous behavior of the conductance.
In the case when the Umklapp interaction creates a large Mott-Hubbard gap 
$2m \gg T_L $   inside the wire, the transport is suppressed  near half-filling 
everywhere inside the gap except for an exponentially small 
region of $V,T < T_L exp(-2m/T_L)$.

\end{abstract}

\pacs{71.10.Pm, 73.23.-b, 73.40.Rw }

%
\multicols{2}
\section{Introduction}

Recent developements in the nano-fabrication technique have made the 
1D interacting electron systems an experimental reality,
and its quantum transport properties have been the subject of extensive
studies both experimentally [1-3] and theoretically [4-16]. 
In realistic experimental set-ups, the quantum wire is attached to 
two-dimensional regions called reservoirs or leads. 
A metallic phase of the infinite wire is known as a Tomonaga-Luttinger
liquid \cite{hald}.
To describe 1D transport phenomena in the realistic configuration a model 
was recently formulated of the inhomogeneous Tomonaga-Luttinger liquid
(ITTL) \cite{mas1,pon,safi}. It 
recovers the conductance $G = 2e^2/h$ observed experimentally
even in the presence of the electron-electron 
interaction in the wire \cite{tar} (below we use the units, where $e=\hbar =1$), although
the previous calculations on an infinitely long wire \cite{kf,of} 
predicted the renormalized conductance
( see \cite{kawa} for the further development).
Calculation of the conductance suppressed by a weak random impurity potential
in this model 
\cite{mas2} had agreed with both the previous theoretical prediction
\cite{of} and an experiment \cite{tar}. 

In this paper we consider effect of opening a spectral gap in the wire on 
the 1D transport. Theory \cite{emery} predicts that in 1D besides band gaps
produced by a periodical potential in the wire at the integer fillings, a 
repulsive interaction between electrons opens Mott-Hubbard gaps at some rational
fillings. Furthermore recently Tarucha et al. \cite{tar2} succeeded in 
introducing the 1D periodic potential with a periodicity of 
order 40nm into the wire 2$\mu$m in length and 
50nm wide. This induces Umklapp scattering. 
The electron density $n$ can be continuously controlled by the gate voltage, 
and one can satisfy the half-filling condition 
within an accessible value of $n$.
If this condition is satisfied the system will becomes a 1D (doped) Mott
insulator with the Mott-Hubbard gap $m$ for the infinite length wire.
Then it will offer an idealistic system to study the 
quantum transport in Mott and doped Mott insulators in 1D.
Earlier a similar experiment had been carried out by Kouwenhoven {\it at al}
\cite{kouw}. Their wire was 3$\mu$m in length, 250-400nm wide and the period of the structure
200nm.  To reduce the impurity backscattering they put the set-up in a
strong magnetic field and observed an interference structure at the integer
edge-state plateaus in  conductance vs. gate voltage. This structure  partly
had been explained as a band gap appearance.

Inspired by these works, we study theoretically 
the $I-V$ characteristic of the wire of length $L$ connected to leads 
at different temperatures $T$
taking into account the Umklapp electron-electron interactions, 
which can be directly compared with the experiments. 
We  formulate our model in the section II and check it on the description of  the band gap
effect at different temperatures. 
We calculate \cite{us} the suppression of the current $ \Delta I$ 
perturbatively in the Umklapp scatterings in the section III. The results are 
summarized in Figs. 1 and 2 for a threshold structure near half filling.
There are two energy scales, i.e., the finite size energy
$T_L = v_c/L$ ($v_c:$ charge velocity) and $E_{thr} \sim v_c \delta n$ 
($\delta n:$ the deviation of the electron density from its value at
the filling $\nu$ equal to $1/2$) 
measuring the incommensurability.
For $T,V>E_c = max(v_c/L,E_{thr})$, the suppression of the 
current $-\Delta I/V$ diverges as $max(T,V)^{4g-3}$  
if the short range
interaction constant $g$ for the forward scattering is less than $3/4$. 
For $T,V<E_c$, on the other hand, the suppression $-\Delta I/V$ goes to zero
as $T,V \rightarrow 0$.
Then we predict the non-monotonous temperature and/or voltage dependence 
of $I$,
which is the clear signature of the Umklapp scattering effect.
For small values of $g $, expected 
when the screening length $\xi_c$ of the interaction determined by the close metallic gates is much larger
than the width of the channel $d$ and $g \propto 1/\sqrt{\ln \xi_c/d} $
\cite{shul}, we predict a few more threshold singularities.
These features could be observed experimentally
by changing the gate voltage, bias voltage, and temperature. 

In the next section IV, motivated by a recent discussion \cite{fuk,mas3} that
in a non-perturbative case of the large Mott-Hubbard gap 
$2m \gg v_c/L \equiv T_L$ the conductance
is strongly suppressed at any low energy (<$m$) inside the gap similar to the band gap 
case,  we consider transport through a 1D Mott-Hubbard insulator of a finite length $L$
beyond perturbative approach.  Our presentation follows \cite{IV} where we used a special value of the 
low energy constant of the interaction to  map the problem onto 
the exactly solvable models.  We find current vs. voltage $V$ at high
temperature $T>max(m,T_L)$ and at low energy $T,V<T_L$.
The result shows 
that for the strong interaction creating a large Mott-Hubbard gap 
$2m \gg T_L $   inside the wire,
the transport is suppressed  near half-filling 
everywhere inside the gap except for an exponentially small 
region of $V,T < T_L exp(-2m/T_L)$

\section{Model}

Our model can be derived following \cite{pon}
from a 1 channel electron Hamiltonian 
\begin{eqnarray}
\lefteqn{ {\cal H}= \int d\!x \{ \sum_\sigma \psi_\sigma^+(x) ( - \frac{
\partial^2_x}{2m^*} - E_F)\psi_\sigma(x)}
\nonumber\\
& & \hspace{10mm} + \varphi(x) \rho^2(x) +
[V_{imp}(x) + V_{period}(x)] \rho(x) \} 
\label{1}
\end{eqnarray}
with the periodic potential $V_{period}(x)$
( period $a$) assumed to be weak enough to justify the perturbative
consideration of the Umklapp backscatterings.
The Fermi momentum $k_F$ and the Fermi energy $ E_F$
is determined by the filling factor $\nu$ as $\nu=k_F a/\pi$ 
and $ E_F \approx v_F k_F$.
In Eq. (\ref{1}) the function $\varphi(x)= const \times \theta (x) \theta (L-x)$ switches on the
electron-electron interaction inside the wire confined in $0<x<L$. This interaction is assumed to be local, as the close metallic gate used in experiments to form the wire inevitably screens the long  range Coulomb.
Contribution of the random impurity potential $V_{imp}(x) \rho(x)$
to the conductance has been considered in \cite{of,mas2}, some results 
of which we will use below. 
 Following Haldane's 
generalized bosonization procedure \cite{hald} to account for the nonlinear 
dispersion one has to write the fermionic fields as 
$\psi_{\sigma}(x)= \sqrt{k_F/(2 \pi)} 
\sum exp\{i(n+1)(k_Fx+\phi_{\sigma}(x)/2) +i\theta_{\sigma}(x)/2 \}$
and the electron density fluctuations as $\rho(x)=\sum \rho_{\sigma}(x), \ 
\rho_{\sigma}(x)=[\partial_x \phi_{\sigma}(x)+2k_F]/(2 \pi) 
\sum exp\{in(k_Fx+\phi_{\sigma}(x)/2)\}$
where summation runs over even $n$ and 
$\phi_{\sigma}, \theta_{\sigma} $ are mutually conjugated bosonic fields 
$[\phi_{\sigma}(x), \theta_{\sigma}(y)]=i 2 \pi sgn(x-y)$. 

After substitution of these expressions into (\ref{1}) and introduction of 
the charge and spin bosonic fields as 
$\phi_{c,s}=(\phi_\uparrow \pm \phi_\downarrow)/ \sqrt{2}$ the Hamiltonian
takes its bose-form ${\cal H}={\cal H}_O + {\cal H}_{bs}$. Here the free 
electron movement modified by the forward scattering interaction is 
described by \cite{mas1,pon,safi}
\begin{eqnarray}
\lefteqn{{\cal H}_O= \int dx \sum_{b=c,s} \frac{v_b}{2}
\{ { 1 \over {g_b(x)} } 
\left({{\partial_x \phi_b(x) } \over {\sqrt{4 \pi}}} \right)^2 }
\nonumber \\
& & \hspace{40mm} + g_b(x) \left({{\partial_x \theta_b(x) }
     \over {\sqrt{4 \pi}}}
\right)^2 \}
\label{2}
\end{eqnarray}
with $g_c(x)=g$ for $x \in [0,L]$ ( $g$ is less than 1 for the repulsive
interaction and it will be assumed below ), $g_c(x)=1$, otherwise and
$v_c(x)=v_F/g_c(x)$. The constants in the spin channel $g_s=1, v_s=v_F$
are fixed by $SU(2)$ symmetry.
Keeping only the most slowly decaying terms among others with the same
transferred momentum one could write the backscattering interaction as
\endmulticols
\vspace{-6mm}\noindent\underline{\hspace{87mm}}
\begin{equation}
{\cal H}_{bs}= {E_F^2 \over v_F} \int_{0}^{L} dx \bigl[
\sum_{even \  m >  0} U_m \cos(2 k_{mF} m x + 
{{m \phi_c(x)} \over \sqrt{2}}) + \sum_{odd \  m >0} U_m
\cos({\phi_s(x) \over \sqrt{2}})
cos(2 k_{mF} m x + {{m \phi_c(x)} \over \sqrt{2}}) \bigr]
\label{3}
\end{equation}
\noindent\hspace{92mm}\underline{\hspace{87mm}}\vspace{-3mm}
\multicols{2}\noindent
A difference in the transferred momentums $2 k_{mF}$ is brought by the
periodical potential with the period $a$:
$k_{mF}=k_F -  \pi l/(m a))$, where $l$ is an integer chosen to minimize 
$|k_{mF}|$. 
We have omitted the $m=0$ term of the first sum: It can contain only the spin field and  cannot affect the current in the lowest perturbative
order.  The most singular is the $m=1$ term of the second sum responsible
for opening the band gap in the infinite wire at $\nu=integer$. The
dimensionless coefficients $U_m$ originate from $\varphi(x)$. 

In the spinless case we should put $\phi_c=\phi_\sigma$ in (\ref{2})
and change the backscattering interaction to
\endmulticols
\vspace{-6mm}\noindent\underline{\hspace{87mm}}
\begin{equation}
{\cal H}_{bs}= {E_F^2 \over v_F} \int_{0}^{L} dx 
\sum_{ m >  0} U_m \cos(2 k_{mF} m x + m \phi_\sigma(x)) 
\label{3'}
\end{equation}
\noindent\hspace{92mm}\underline{\hspace{87mm}}\vspace{-3mm}
\multicols{2}\noindent
To generalize our perturbative results of the part III to the spinless case 
one just needs to transform $m, k_{mF}$ to $\sqrt{2}m, k_{mF}/\sqrt{2}$ in the expressions
written in the spin case for even $m$ and take $m$ arbitrary integer.

It is instructive first to examine how the band gap shows out in
transport properties of the wire filled with the non-interacting 
electrons. The model is equivalent to a Dirac equation with the
mass switched on inside the wire:
\endmulticols
\vspace{-6mm}\noindent\underline{\hspace{87mm}}
\begin{equation}
{\cal H}_{D}=i\sum_{a=R,L=\pm} \psi^+_a ( \mp v_c \partial_x) \psi_a
-m \varphi(x) [e^{i2k_Fx}\psi^+_L(x)\psi_R(x)+h.c.]
\label{24}
\end{equation}
\noindent\hspace{92mm}\underline{\hspace{87mm}}\vspace{-3mm}
\multicols{2}\noindent
with $m=\pi E_F U_1/2$ in the spin case of Eq.  (\ref{3}) and  $m=\pi E_F U_1$ in the spinless case of Eq. (\ref{3'}). 
Here $\psi_{R(L)}$ describes the right (left) chiral electron and 
$k_F$ is counted from $\pi l/a$ at $\nu \simeq l$.
The transport is determined by the transmittance $D$ which is a function of
the electron energy $\varepsilon$ counted from the one of the middle of the gap $v_F \pi l/a$ : 
\begin{equation}
D(\varepsilon)=\left[1+ m^2{\sin^2(\sqrt{\varepsilon^2-m^2}t_L) \over \varepsilon^2-m^2}\right]^{-1}
\label{25}
\end{equation}
where the analytical extension is assumed at $\varepsilon < m$ and $t_L=L/v_F$.
In particular, the linear bias conductance $G$ is equal to
\begin{equation}
G(T, \mu_0)={1 \over 2 \pi T}\int d \varepsilon  \frac{D(\varepsilon)}{1+\cosh((\varepsilon - \mu_0)/T)}
\label{26}
\end{equation}
for spin electrons where $\mu_0=k_F v_F$ gives a deviation of the chemical potential of the wire from that of
the $\nu=l$ filling.  The conductance has two regimes of behavior.

{\it 1. High temperatures} $T > T_L, m$ - The asymptotics to (\ref{26}) can be written as
\begin{equation}
G={1 \over \pi} \left( 1- {m \over 2 T}
\frac{B(m / T_L)}
{1+\cosh(\mu_0/T)} \right)
\label{27}
\end{equation}
after noticing that $D$ is a quickly oscillating function against the slowly varying temperature factor.
The coefficient $B$ changes slowly from 0 to $\simeq \pi$ and is defined below in Eq. (\ref{45}).
The expression (\ref{27}) shows that at the high temperature the band gap produces a smooth well in the conductance vs. chemical potential which becomes more narrow and deeper as $T$ decreases.
   
{\it 2. Low temperatures} $T < T_L$- The conductance is about $D(\mu_0)/\pi $. It is suppressed in the middle of the gap $ G(0)=\left[\pi (1+ (mt_L)^2) \right]^{-1}$
and approach its maximum $1/\pi$ away from the gap oscillating with period $\pi/t_L $. In the
limit $m t_L \ll 1$ this interference structure is perfectly periodical, as 
\begin{equation}
 \Delta G_1=-(m t_L)^2 \frac{\sin^2((k_F-\pi l/a)L)}{ ((k_F-\pi l/a)L)^2}
\label{28}
\end{equation}
 Therefore the conductance has $L/a$ humps in between two neighbor band gaps.  We expect this 
being correct  at any finite $m$ from comparison with a tight binding model. In our model it holds on asymptotically if $\pi v_c/(a m) \gg 1$. Then the model works well.

\section{Narrow gap: perturbative approach}

In this section we assume that the dimensionless coefficients $U_m$ in Eq.(\ref{3}) are
small enough to justify perturbative calculation of the current. The variation
of the current due to the backscattering is given by
: $\Delta\! I=-i/(2 \sqrt{2} \pi) \int d\!x \, [\partial_x \theta_c(x), 
{\cal H}]= \sqrt{2} \int d\!x (\delta/\delta\!\phi_c(x)) {\cal H}_{bs}$. 
At finite voltage $V$ applied symmetrically to neglect the momentum
transfer variation, the average of $\Delta\! I$ 
decomposes into sum of the different backscattering mechanism contributions
$<\Delta \! I_m>$ in the lowest order.
The even $m$ terms involving only $\phi_c$ field are equal to 
%
\endmulticols
\vspace{-7mm}\noindent\underline{\hspace{87mm}}
\begin{eqnarray}
<\Delta\! I_m>= -{m \over 4} \bigl({{U_m E_F^2} \over v_F}\bigr)^2
\int_{- \infty}^{\infty}dt \int\! \int_{0}^{L} d \!x_1 d \! x_2
<e^{im \phi_c(x_1,t)/\sqrt{2}} e^{-im \phi_c(x_2,0)/\sqrt{2}}>
\bigl[e^{im(2k_{mF}(x_1-x_2)+Vt)} - h.c. \bigr].
\label{4}
\end{eqnarray}
\noindent\hspace{92mm}\underline{\hspace{87mm}}\vspace{-3mm}
\multicols{2}\noindent
The current operator $\Delta \! I_m$ has a high energy scaling
dimension $m^2 g/2$ and a free electron ($g=1$) behavior at low energy.
We will see below that the integral (\ref{4}) scales at low energy 
with $(m^2-1)$ exponent and with $(m^2 g-2)$ exponent at high energy.
The most singular behavior is
due to Umklapp backscattering at $m=2$ with the threshold voltage 
$V=E_{thr}=2k_{2F} v_c$ going to zero at the half filling. We assume $V,E_{thr}\ge 0$ below.
Less singular
correction with $m=4$ could become relevant at the one and three quarters fillings
 and so on.
Expressions for the odd $m$ terms include additionally a spin field
correlator $<e^{i \phi_s(x_1,t)/\sqrt{2}} e^{-i \phi_s(x_2,0)/\sqrt{2}}>$
under the integrals in (\ref{4}). 
The high energy dimension of $\Delta \! I_m$ in this case is $m^2 g/2+1/2$.
The most singular behavior occurs to
the $m=3$ term at the one and two thirds fillings. It has two threshold energies
$E_{thr\,c,s}=2 k_{3F} v_{c,s}$ for $v_c \neq v_s$.  Neglecting a change of the TLL compressibility produced by the Umklapp scattering we can relate \cite{us2} the threshold energy to a deviation of the chemical potential of the wire $\mu$ at $V=0$  from that of the rational filling. Since $\Delta \rho_c=2 k_F/\pi$ and $ \partial \mu/\partial\rho_c=\pi v_c/(2g)$
we gather $E_{thr,c}=2g\mu$. However, in an experiment it is the average 
of the electrochemical potentials of the leads but not $\mu$ that is known.
The latter is proportional to $\mu$ with the coefficient $[1+2ge^2/(\pi v_c \hbar c_g)]$
if the gate voltage is fixed \cite{us2}. This coefficient is about $1$ if the density $c_g$
of the capacitance between the wire and the screening gate is large.

Correlator of the charge field exponents $e^{i \phi_c(x,t)}$,
evolution of which is specified by ${\cal H}_O$, 
could be compiled from the correlators of the uniform TL
liquid $K(x,t) = K(x,t,g,v_c)$ ($K(x,t,g,v)\equiv  
(\alpha \pi /\beta )^{2g} /( \prod_{\pm} 
\sinh^g(\pi (x/v \pm (t-i \alpha))/\beta )) )
$ in the following way \cite{mas1,mas2} 
\endmulticols
\vspace{-7mm}\noindent\underline{\hspace{87mm}}
%
\begin{eqnarray}
\lefteqn{
<e^{i \phi_c(x,t)} e^{-i \phi_c(y,0)}>=K(x-y,t) \prod_{\pm, n=1}^{\infty}
\bigl( { { K(2nL,0)} \over {K(2nL \pm \mid x-y \mid ,t)} } \bigr)^{-r^{2n}}}
\label{5} \\
& & \hspace{15mm} \times \prod_{ n=0}^{\infty} \bigl(
 { { K(2(nL+x),0) K(2(nL+y),0)} \over {K^2(2nL+x+y,t)} }\bigr)^{-r^{2n+1}/2}
\prod_{ n=1}^{\infty} \bigl(
 { { K(2(nL-x),0) K(2(nL-y),0)} \over {K^2(2nL-x-y,t)} }\bigr)^{-r^{2n-1}/2}, 
\nonumber
\end{eqnarray}
\noindent\hspace{92mm}\underline{\hspace{87mm}}\vspace{-3mm}
\multicols{2}\noindent
Here $\beta$ is inverse temperature $1/T$ and $\alpha =1/E_F$ is the 
ultraviolet cut-off. This complicated form comes about through a multiple
scattering at the points of joint $x=0,L$. As a result of the scattering the 
correlator $<\phi_c(x,t) \phi_c(y,0)>$ becomes an infinite sum of the
uniform correlators taken along the different paths connecting points
x and y and undergoing reflections from the boundaries at $x=0,L$. Each
reflection brings additional factor $r=(1-g)/(1+g)$. 
The similar correlator
$<e^{i \phi_s(x,t)} e^{-i \phi_s(y,0)}>$ for spin field is
$K(x-y,t,1,v_s)$.
Below we analyze the current
corrections (\ref{4}) for high ($T>1/t_L=T_L$) and low
($T \ll T_L$) temperatures, respectively. 

\noindent
1.{\it High temperatures} $T>T_L$ - The uniform correlator $K(x,t)$ goes down
exponentially if distance between the points $\mid x \mid$ exceeds
the inverse temperature. Therefore only paths with length less than 
$\beta $ contribute to the correlator (\ref{5}). This means that the 
high temperature form of the correlator (\ref{5}) reduces to the first
multiplier $K(x-y,t)$ up to a factor $(1+O(exp(-t_L/\beta ))$.
Neglecting $O(T_L/T)$
quantity we can extend integration over $x_1-x_2$ in (\ref{4}) from $-\infty $
to $+\infty $. Then calculation of the $m=2$  contribution
reduces to finding  Fourier transformation $F_{2g}(q,\varepsilon)$ of 
the correlator $K^2(x,t,g, v_c)$:
\endmulticols
\vspace{-7mm}\noindent\underline{\hspace{87mm}}
%
\begin{eqnarray}
<\Delta\! I_2>={1 \over 4} \left({{U_2 E_F^2} \over g}\right)^2
t_L \sum_{\pm} \mp F_{2g}(2 E_{thr},\pm 2V)=
-2 \bigl({{2^{2(g-1)}U_2} \over {\Gamma(2g) g}}\bigr)^2 {E_F^2 \over T_L}
\bigl({{\pi T}\over E_F} \bigr)^{4g-2} \sinh({V \over T})
\prod_{\pm,\pm} \Gamma(g \pm i{{V \pm E_{thr}}\over {2 \pi T}})
\label{8}
\end{eqnarray}
One can easily see its behavior making use of the following asymtpotics:
\begin{eqnarray}
<\Delta\! I_2> \propto -\left({U_2  \over g}\right)^2 {E_F^2 \over T_L}
\left\{ \matrix{((V^2-E_{thr}^2)/E^2_F)^{2g-1}, 
&  V \gg E_{thr},\ T \cr
((V+E_{thr}) T/E^2_F)^{2g-1},\ \  &  V \approx E_{thr} \gg T \cr } 
\right.  \label{9} \\
<\Delta\! I_2> \propto -\left({U_2  \over g}\right)^2 {E_F^2 \over T_L}
\sinh{V \over T } \left\{ 
\matrix{ e^{-E_{thr}/T}((E_{thr}^2-V^2)/E^2_F)^{2g-1}, \ \ 
& E_{thr} \gg V,\ T \cr
(T/E_F)^{4g-2}, &  V \ , E_{thr} \ll T  \cr } \right.
\nonumber
\end{eqnarray}
\noindent\hspace{92mm}\underline{\hspace{87mm}}\vspace{-3mm}
\multicols{2}\noindent
\begin{figure}[htbp]
\begin{center}
\leavevmode
\psfig{file=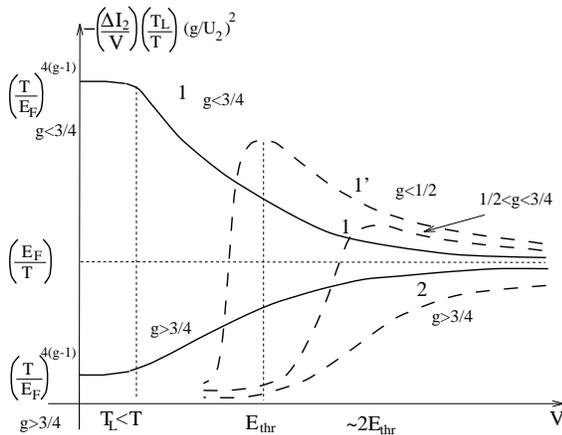,width=2.9in,angle=270}
\narrowtext{ \caption{
           Schematic voltage dependence of the high temperature
            current corrections produced by the $m=2$ Umklapp 
            interaction $\Delta I_2$. Solid lines, $E_{thr}=0$; 
            dashed lines, $E_{thr}\gg T$.  \label{cty}
           }}
\end{center}
\end{figure}
These asymptotics show that the threshold singularity in the current
voltage dependence diverges as $((V-E_{thr})/T)^{2g-1}$ if $g<1/2$
(Fig.\ref{cty}). 
It becomes stronger in the differential conductance
dependence. At $g>1/2$ the differential 
conductance correction $d\! G_2$ behaves as $-(V/E_F)^{4g-3}$, and 
saturates at $-(T/E_F)^{4g-3}$ below $T$, if $E_{thr}<T$; otherwise,
the correction shows divergence $-((V-E_{thr})/T)^{2g-2}$
smeared over $T$ scale near the threshold and 
becomes suppressed exponentially as $-exp(-E_{thr}/T)$ below it.
Eq. (\ref{9}) predicts also that the linear bias conductance $G$ as a function of $E_{thr}$ proportional to $\mu$ and the gate voltage  
has a $T$ wide well at half filling similar to the band gap case discussed in the section II. However,
the depth of the well $\propto T^{4g-3}$ increases with decrease of  $T$ only if the repulsive interaction is strong enough. This is a bit unexpected result since we know \cite{emery} that any repulsive interaction opens the 
spectral gap at half-filling in the infinite wire. It will be discussed further 
in the next section. 

Generalization to the other even $m$ expressions for the backscattering current of the spin electrons
needs just changing:
$4g \rightarrow m^2 g \ , 4k_{2F} \pm 2V \rightarrow m(2k_{mF} \pm V)$. In the case of 
the spinless electrons: $4g \rightarrow 2m^2 g \ , 4k_{2F} \pm 2V \rightarrow m(2k_{mF} \pm V)$ with 
arbitrary integer $m$. In particular, we find again $T^{-1}$-dependence for decrease of the conductance produced by the band gap opening in the spectrum of free electrons.
The edge singularity is characterized
by a half of the scaling dimension for $\Delta I_m$ since only
one chiral component of the field $\phi_c$ contributes.

As to the odd $m$ terms, 
the two threshold energies
$E_{thr\,c,s}=2 k_{mF} v_{c,s}$ become distinguishable if their difference
exceeds $T$.
The leading high-temperature current correction reads as:
\endmulticols
\widetext
\vspace{-7mm}\noindent\underline{\hspace{87mm}}
\begin{equation}
<\Delta\! I_m>={ m \over 4} \left({{U_m E_F^2} \over {2 \pi g}}\right)^2
t_L v_s 
\sum_\pm \mp \int \! \int d \! q d \! \varepsilon
F_{\frac{1}{2}}(mE_{thr\,s}-q v_s, \pm mV-\varepsilon)
F_{{m^2g} \over 2}(q v_c,\varepsilon)
\label{10}
\end{equation}
\noindent\hspace{92mm}\underline{\hspace{87mm}}\vspace{-3mm}
\multicols{2}\noindent
Substituting zero temperature form of $F_a$ function
$F_a(q,w)=8[sin(\pi a) \Gamma(1-a)]^2 (\alpha /2)^{2a}\prod_\pm
(w\pm q)^{a-1} \theta(w\pm q)$ in (\ref{10})
one can gather that the current correction 
behaves as $-(V/E_F)^{m^2g-1}$ at large voltage $V > E_{thr\,c,s}$, has a 
leading singularity $-((V-E_{thr\,c})/T)^{m^2g/2}$ smeared over $T$ scale
near the first threshold and $-((V-E_{thr\,s})/T)^{m^2g-1/2}$ near the second
threshold (we assume $v_c>v_s$). Below the lowest threshold it becomes
exponentially suppressed. These singularities result in the divergences
of the differential conductance or higher derivatives of the current
in voltage.
The threshold behavior of the $m=3$ term of the differential
conductance correction is
divergent at $E_{thr\ c}$ if $g<2/9$ and at $E_{thr\ s}$ if $g<1/6$. On the other hand,
it is unlikely that the splitting of the threshold energies could be observed at the integer filling
factor where opening the band gap eventually makes electrons non-interacting.  

\noindent
2.{\it Low temperatures} $T,V<<T_L$ - With lowering temperature 
we should expect that above current 
correction dependencies will be modulated by a $\pi T_L$ quasiperiodical
interference structure \cite{pon2,naz} and  also 
a new low energy scaling behavior 
of the current correction operators  will appear at $V,T<T_L$.
The dominant contribution to the integral of (\ref{4})
comes from long times $t \gg t_L$.  One can neglect the spacious dependence
compared with large $t$ in (\ref{5}) and keep the multipliers with the number of 
reflections $n<n^*=\beta/(2 t_L \pi )$ only to come to the long time 
asymptotics:
\endmulticols
\vspace{-7mm}\noindent\underline{\hspace{87mm}}
\begin{equation}
<e^{i \phi_c(x,t)} e^{-i \phi_c(y,0)}>=e^{\gamma(T_L/T)} \bigl( {\alpha
\over t_L } \bigr)^{2g} \bigl({{(\pi t_L/\beta )^2} \over {
sinh(\pi (t-i\alpha)/\beta) sinh(\pi (-t+i\alpha)/\beta)}} \bigr)^{1-z}
\bigl({\sqrt{x y (L-x)(L-y)} \over {L^2}} \bigr)^{2rg}
\label{6} 
\end{equation}
\noindent\hspace{92mm}\underline{\hspace{87mm}}\vspace{-3mm}
\multicols{2}\noindent
where $z(T_L/T)=r^{\beta/(t_L \pi)}$ and $\gamma(T_L/T)$ approach the constant
 $\gamma(\infty)$ on the order of 1 as $\ln (T_L/T) z(T_L/T)$.
Our asymptotic analysis following in essential 
Maslov's paper \cite{mas2} shows that the low energy exponents 
approach their free electron values as $ exp[T_L \ln r/(T \pi)] $.
 The effect accounts for prolongation
of the paths due to the finite reflection. 
In particular, it determines the coefficient $c(g)$ of the $T^2$
corrections to the non-universal zero temperature value of the 
conductance  variation due to impurities:
$\Delta G_{imp} \propto - (L/l) (T/E_F)^{g-1}
(1 - c(g) (T/T_L)^2)$ 
in a universal way \cite{us3}. 
After substitution of (\ref{6}) into Eq. (\ref{4}),  the current suppression
produced by the even $m$ terms of the  interaction  
becomes equal to:
\endmulticols
\vspace{-7mm}\noindent\underline{\hspace{87mm}}
\begin{equation}
<\Delta\! I_m>=-{{m 2^{m^2(1-z)} e^{m^2\gamma/2 }}\over {\Gamma(m^2(1-z))}} 
\left({{U_m } \over g}\right)^2 R_{{m^2g}\over 2}(2m k_{mF}L) T_L 
\left({{\pi T } \over T_L}\right)^{m^2-1} 
\left({T_L \over E_F}\right)^{m^2(g-1)} f_{{m^2}\over 2}(V/T)
\label{11}
\end{equation}
where function 
$f_{a}(x)=\sinh(x) \prod_\pm \Gamma(a \pm ix/\pi)$
characterizes the $V-T$ cross over. 
It approaches  
$ \Gamma^2 (a)x(1 - (\ln\Gamma(a))'\!' (x/\pi)^2)$ at $ x \ll 1 $
and $\pi (x/\pi)^{2a-1}$ at $x \gg 1 $.
Function $R$ specifies the $k_F - 1/L$ crossover as:
\begin{eqnarray}
R_{2g}(x)= {{\pi \Gamma^2(1+2rg)} \over x^{1+4rg}} J^2_{1/2+2rg}(x/2) 
\simeq \Gamma^2(1+2rg)
\left\{ \matrix{ \pi/(4^{1+4rg} \Gamma^2(3/2+2rg)), \ \ 
&  x \ll 1 \cr
4 sin^2(x/2-\pi r g)x^{-2-4rg},\ \  & x \gg 1  \cr } 
\right. 
\label{12} 
\end{eqnarray}
\noindent\hspace{92mm}\underline{\hspace{87mm}}\vspace{-3mm}
\multicols{2}\noindent
It brings out an interference structure in  the conductance versus the chemical potential at low energy. This
structure coincides with the one of Eq. (\ref{28}) at $m=1$.  At larger $m$, however, the oscillations are 
more frequent. In particular, there can be the unchanged $L/a$ number of maximums of the conductance in
between its neighbor minimums at the half-filling and the integer filling. This effect has not been accounted for in \cite{kouw}. 
\begin{figure}[htbp]
\begin{center}
\leavevmode
\psfig{file=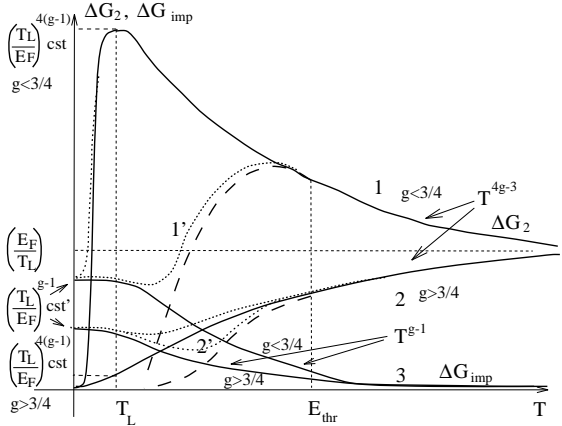,width=2.9in,angle=270}
\narrowtext{
  \caption{Schematic temperature dependence of the conductance 
         corrections produced by the $m=2$ Umklapp interaction 
         $\Delta G_2$ at $g<3/4$ (lines1)
         and at $g>3/4$ (lines2) and by the random impurity 
         potential $\Delta G_{imp}$ \cite{of,mas2} 
         (lines 3).  
         Solid lines, $E_{thr}=0$; dashed lines, $E_{thr}\gg T_L$.
         The dot lines 1 and 2 (1' and 2') are the full conductance 
         correction at $E_{thr}=0$ (finite $E_{thr})$.   \label{cn}
        }}
\end{center}
\end{figure}

The odd $m$ terms of the spin electron current will meet Eq.(\ref{11}) after substitution
$m^2+1$ instead of $m^2$ into the powers and the index of the $f$ function in
this equation. In the spinless case we have to
substitute $2m^2$ there and into the index of $R$. Combining the  
above results we can outline a temperature 
dependence of the conductance 
correction produced by the $m=2$ Umklapp interaction (Fig.\ref{cn}).
For spin electrons its magnitude increases/decreases
following $(E_F/T_L)(T/E_F)^{4g-3}$ as $T$ going down above $T_L$ 
and follows $(T/T_L)^{2} (T_L/E_F)^{4g-4}$, if  
$E_{thr}<T_L$; otherwise,
the correction starts to decrease exponentially  $exp(-E_{thr}/T)$
below $E_{thr}$ and keeps on decreasing like 
$(T/T_L)^2 (T_L/E_F)^{4g-4} (T_L/E_{thr})^{2+4rg} 
\sin^2(2k_{2F} L-\pi rg)$
 below $T_L$. The $T(>T_L)$ dependence is similar to that of the 
{\it conductivity} of infinite wire found by Giamarchi \cite{g}.
 Similar dependence with $T$ replaced by $V$ could be
predicted for the zero temperature differential conductance $d G_2(V)$ 
at $V<T_L$. 

In summary under perturbative condition
we have described a hierarchy of the threshold features
produced by the Umklapp backscatterings at
the rational values of the occupation number inside the 1D channel
connecting two Fermi liquid reservoirs. In the differential
conductance (its derivative) vs. the chemical potential /threshold energy at a finite voltage, 
the threshold structure is an asymmetric peak of width $ max(T,T_L)$ 
located at the crossover $E_{thr}\approx V$ as the chemical potential moves away from the rational filling. These peaks are produced by {\it any} repulsive interaction at the half-filling of spin electrons in the wire. In the conductance vs. temperature,
we predicted a maximum below $E_{thr}$ due to crossover from the 
Umklapp backscattering to the impurity suppression and an asymmetric 
minimum at $E_{thr}$ if the interaction is strong enough.  However, if the interaction
is weak so that $g>3/4$ the suppression of the conductance even at the half-filling may be difficult for
observation while $m/T_L  \ll  1$.

\section{1D Mott-Hubbard insulator: non-perturbative results. }

In this section, to clear up the difference between the Mott-Hubbard insulator 
and the band gap one, we map the problem at low energies and at high 
temperature onto the exactly solvable models making use of a free fermion
value of the constant $g$ of the forward scattering inside the wire. 
The results are shown in Figs. 3 and 4. At low energies when 
$T,V \ll T_L$ 
($T$: temperature; $V$: voltage), we have found that  
a new energy scale $T_x \propto T_L exp[-2\sqrt{m^2-\mu^2}/T_L]$ 
appears in the system 
if the chemical potential $\mu =E_{thr}/(2g)$ of the wire is small enough:  
$\sqrt{m^2-\mu^2}/T_L \gg 1$. 
Below $T_x$ the conductance is not suppressed 
and the current increases linearly.
Above this energy the current saturates and the conductance goes down as $T_x/T$
reaching small values $\approx exp[-2\sqrt{m^2-\mu^2}/T_L]$ at $T \approx T_L$. 
At high temperature
$T \gg T_L,m$ we confirmed the asymptotical behavior of the conductance: 
$G=(1-cst {m \over T} (1+\cosh{\mu \over T})^{-1} )/\pi$  for a
Mott-Hubbard insulator which has been found in the previous section in the perturbative regime of a small gap \cite{us}.
A brief physical explanation to these results follows. 
At low energies $T < T_L$ and $\mu \ll m$ the charge field is quantized
inside the wire at its values related to the degenerate
sin-Gordon vacua. Rare low energy excitations tunnel through the wire
with the amplitude $\propto exp[-m/T_L]$ as (anti)solitons
switching the quantized value of the field. 
The whole process of tunneling, however, includes
transformation of the reservoir electron into the sin-Gordon quasiparticles
and back. This transformation results in a non-trivial scaling dimension
of the tunneling operator equal to  $1/2$ for the Mott-Hubbard insulator 
connected to the Fermi liquid reservoirs
independently of any parameters. 
In the case of the band insulator, this dimension is marginal ($=1$): 
the transformation is trivial and does not introduce additional
energy dependence.
The infrared relevantness of the tunneling with the $1/2$ dimension brings
out above resonance at zero energy. Meanwhile, the exponentially small tunneling
amplitude specifies the narrow width of  this resonance equal to
the crossover energy. Increase of $|\mu |$ favors tunneling of the
quasiparticles of the same sort 
and ultimately produces their finite density in the wire.
Then the interaction between these quasiparticles described with
the two-particle $S$-matrices \cite{zam} dependent on $g$ emerges. At low momenta
the $S$-matrix for the quasiparticles of the same sort is inevitably free fermion
like, as at $g=1/2$. 
It manifests in the renormalization group (RG) flow derived from the 
Bethe anzats solution for the massive phase \cite{ktrg}  of the sin-Gordon model
and in the exponent calculated for the Tomonaga Luttinger liquid
(TLL) phase at low density \cite{onehalf}. Increase of $T$, on the other
hand, is expected to entail, first, a thermally activated behavior of the
conductance $\propto exp[-2m/T]$ at $T_L<T<m$ \cite{rice} and then a
power law dependence at $m<T$. Since the effective value of $g$, in general,
scales with energy,
the $1/T$ dependence we found for
$g=1/2$ may vary at higher energies $T \gg m$ depending on the high energy value 
of $g$.

Transport through the finite length wire under a constant voltage $V$ between
the left and right leads could be described in the inhomogeneous
Tomonaga-Luttinger liquid model (TLL) with the Lagrangian \cite{us,us2}:
$\int dx \{ \sum_b {\cal L}_b(x,\phi_b,\partial_t \phi_b) + 
{\cal L}_{bs}(x,Vt,\phi_c, \phi_s) \}$ associated to the Hamiltonian (\ref{2},\ref{3}). The 
bosonic fields $\phi_b(x,t), b=c,s$  relate to the deviations of 
the charge and
spin densities from their average values as following: 
$\rho_{b}(x,t)=(\partial_x \phi_{b}(x,t))/(\sqrt{2} \pi)$,
respectively. The first part of the Lagrangian describes a
free electron motion modified by the forward scattering interaction.
The second part of the Lagrangian
introduces backscattering inside the wire.
Only its term corresponding to the Umklapp process of four
Fermi momenta transfer 
is important
near half-filling. This term does not involve
the spin field. Therefore, our consideration will be restricted to the charge 
field only. For the clean wire this field is characterized by the Lagrangian:
\endmulticols
\widetext
\vspace{-6mm}\noindent\underline{\hspace{87mm}}
\begin{equation}
\int d x {\cal L}_t= \int dx \bigl[ \frac{v_c(x)}{2g(x)}
\{ 
{1 \over v_c^2} \left({{\partial_t \phi_c(t,x) }
     \over {\sqrt{4 \pi}}}
\right)^2 -
\left({{\partial_x \phi_c(t,x) } \over {\sqrt{4 \pi}}} \right)^2 
\}
-
{E_F^2 U \over v_F} \varphi(x) \cos(4 k_{2F} x + 2 V t +
\sqrt{2} \phi_c(t,x)) \bigr]
\label{41}
\end{equation}
\noindent\hspace{92mm}\underline{\hspace{87mm}}\vspace{-3mm}
\multicols{2}
\noindent
where $\varphi(x)= \theta (x) \theta (L-x)$ specifies a one channel wire
of the length $L$ adiabatically attached to the leads $x>L,x<0$
and $v_F(E_F)$ denotes the Fermi velocity(energy) in the channel. 
The parameter $4k_{2F}$ varies the chemical potential $\mu$ 
of the wire from its zero value at half-filling. In a real experiment as we discussed before this chemical potential is linearly changed by variation of the electrochemical potential of the screening gate or the average of the reservoir potentials.  Outside the
Hubbard gap this parameter coincides with the momentum
transferred by the backscattering: four Fermi momenta minus a vector of 
the reciprocal lattice, and relates the present results to the 
ones \cite{us} of the previous section. We assume $\mu \ge 0$ below.
The constant of the forward scattering varies from 
$g_c(x)=g$  inside the wire  ($x \in [0,L]$) to 
$g_c(x) \equiv g_\infty =1$  inside the leads, and the Umklapp scattering of 
the strength $U$ is introduced inside the wire. The charge velocity
$v_c(x)$ changes from $v_F$ outside the wire to a some constant $v_c$
inside it. In the absence of the Umklapp scattering, $v_c \simeq v_F/g$ and $0<g<1$
is determined by the forward scattering amplitude of the bare short range 
interaction between electrons. Approaching the half-filling put the
Umklapp scattering on. It entails an essential renormalization of the 
low energy value of $g$, which flows to its free fermion value $g=1/2$ 
in the massive phase \cite{ktrg} $(\mu < m)$ 
where the coefficient of the $\cos $-term scales to $\simeq m^2$
and on approaching 
this phase \cite{onehalf} $|\mu | \searrow m$. 
This value of $g$ will be assumed below.
The zero frequency current through 
the wire equals $I=V/\pi + <\hat{I}_{bsc}>$, where the backscattering current 
\cite{us2} is
$\hat{I}_{bsc}=-2 E_F^2 U/v_F \int_0^L dx \sin(4k_F x +2Vt +\sqrt{2} \phi_c(x))$.
It will be shown later that $2 \pi E_F U$ is a half gap $m$, 
opened by the backscattering (\ref{1}) in the charge mode spectrum inside the wire.

\noindent
1.{\it High temperatures} $T>T_L,m$ - 
The average backscattering current 
$<\hat{I}_{bsc}>=\int D\phi \hat{I}_{bsc} exp\{i\int dt \int dx 
({\cal L}_c+{\cal L}_{bsc})\}$ can be written as a formal infinite series
in $U$. Each term of it is an integral of product of the free bosonic 
correlators $<exp\{ i \sqrt{2}\phi(x,t)\}exp\{ -i \sqrt{2}\phi(y,0)\}>$. 
Such a correlator approaches its uniform TLL expression when 
$x,L-x,y,L-y \gg v_c/T$. Substitution of this form into the above series 
allowed us \cite{us} to find $L$-proportional part of the backscattering 
current neglecting the boundary contribution in the perturbative case.
However, the problem is not perturbative, in general, due to a finite
gap $2m$ creation. Therefore, application of the uniform correlator
will give us a part of the backscattering current $\propto min(L,v_c/m)$
with the relative error $O(max(T_L,m)/T)$, which is of the order of 
ratio
of the border piece $\propto v_c/T$ to the essential part of 
the "bulk" one. 
This relates to the high-temperature asymptotics of the whole current.
\begin{figure}[htb]
\begin{center}
\leavevmode
\psfig{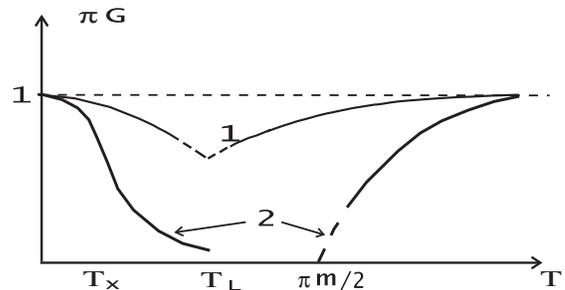}
\narrowtext{ \caption{
           Schematic linear bias conductance $G$ vs.
            temperature near half-filling $\mu \ll T_L$: 
            curve 1 for the weak interaction $m \ll T_L$,
            curve 2 for the strong one.  \label{cty}
           }}
\end{center}
\end{figure}
Calculation of above series with the uniform TLL correlator 
is equivalent to expanding the value
$g=1/2$ into the leads. 
Following Luther and Emery \cite{le}
we map this bosonic Lagrangian 
onto the free massive fermion one \cite{fuk,mas3} with the density
of Lagrangian
\endmulticols
\widetext
\vspace{-6mm}\noindent\underline{\hspace{87mm}}
\begin{equation}
{\cal L}_{F}=i\sum_{a=R,L=\pm} \psi^+_a (\partial_t \pm v_c \partial_x) \psi_a
-m \varphi(x) [e^{i(4k_Fx+2Vt)}\psi^+_L(x,t)\psi_R(x,t)+h.c.]
\label{42}
\end{equation}
\noindent\hspace{92mm}\underline{\hspace{87mm}}\vspace{-3mm}
\multicols{2}
\nopagebreak
Here $\psi_{R(L)}$ is right (left) chiral fermion field. The fermionized
backscattering current $\hat{I}_{bsc}=
2mi \int_0^L dx [exp\{i(4k_Fx+2Vt)\}\psi^+_L(x,t)\psi_R(x,t)-h.c.]$ is the
doubled backscattering current for the fermions \cite{us2} under 
doubled voltage. To find its average we just need to know 
the fermionic reflection coefficient $R$
as a function of dimensionless energy $\omega = \varepsilon/m$:
\begin{equation}
R(\omega)=\frac{\sin^2(\sqrt{\omega^2-1}\bar{t}_L)}
{(\omega^2-1)+\sin^2(\sqrt{\omega^2-1}\bar{t}_L)}
\label{43}
\end{equation}
where $\bar{t}_L\equiv m/T_L$ denotes the dimensionless traversal time. 
The analytical continuation is assumed for $|\omega | < 1$.
Since the chemical potential for the right/left chiral fermions
is $\mu\pm V$, respectively,
the total current can be expressed as
\begin{eqnarray}
\lefteqn{I={V \over \pi} - {m \sinh(V/T) \over \pi} \times }
\nonumber \\
& & \hspace{5mm} \int d \omega \frac{R(\omega)}
{\cosh((m \omega-\mu)/T)+ \cosh(V/T)}
\label{44}
\end{eqnarray}
Only the leading term in $max(m,T_L)/T$ of the right hand side of (\ref{44})
is meaningful. 
Extracting it, we find the high-temperature asymptotics as following
\begin{eqnarray}
\lefteqn{
I={V \over \pi} - {m\over \pi}
\frac{ \sinh(V/T) B(mt_L)}{\cosh((\mu)/T)+ \cosh(V/T)} }
\label{45}\\
& & \hspace{3mm} B(x)\equiv \int d \omega 
\frac{\sin^2(\sqrt{\omega^2-1}x)}
{(\omega^2-1)+\sin^2(\sqrt{\omega^2-1}x)}
\nonumber
\end{eqnarray}
where function $B(x)$ increases as $\pi x$ at small $x>0$ and approaches 
the constant $\simeq \pi$ at $x\gg 1$. Accuracy of this calculation
of (\ref{44}) 
may be written as a factor $1+O(max(m,T_L)/max(T,V))$ to (\ref{5}) if 
$|\mu| \ll max(T,V)$ or as 
$1+O([max(m,T_L)/max(T,V)](max(T,V)/\mu)^2e^{|\mu|/T})$, otherwise.
The high-temperature conductance 
(Fig.\ref{cty})
\begin{equation}
G={1 \over \pi} \left( 1- {m \over T}
\frac{B(m / T_L)}
{1+\cosh(\mu/T)} \right)
\label{46}
\end{equation}
approaches zero at $T \approx m$ if the gap is large enough $m/T_L \gg 1$
and $|\mu|<m$. This asymptotics gives an adequate description of  the whole high 
temperature region $T > m$ while a high energy value of $g$ was about 1/2. Then it
complies with the high temperature conductance (\ref{9})  we have found in the previous section.
At others values of $g$ we would expect the conductance changing its behavior from (\ref{9}) to (\ref{46}) as $g$ is scaling with lowering the temperature. This crossover could result in a non-
monotonous  dependence of the conductance on the temperature, since at high energy $g>3/4$ we have 
seen that the conductance is increasing with decrease of the temperature.

\noindent
2.{\it Low energies} $T,V<<T_L$ - To find a low energy model for our 
problem we have to integrate out all high energy modes. We will try to escape
direct integration following
Wiegmann's effective way of constructing the Bethe-ansatz solvable 
models \cite{w} for the Kondo problem and for the screening of a resonant 
level  \cite{w,p}. First, let us substitute $\phi/\sqrt{2} $
instead of $\phi$ in (\ref{41}). It makes fermions interacting inside the
leads and non-interacting inside the wire. Their passage through the 
wire may be described with the one-electron $S$-matrix dependent on the
electron momentum. The interaction between electrons in the leads 
with some two-particle  $S$-matrix. Then the solution could be constructed 
if the 
proper commutation relations between the $S$-matrices are met. Being 
interested in the variation of energy less then $T_L$ around the Fermi level, 
one can simplify the solution keeping the one-particle 
$S$-matrix constant equal to its value on the Fermi level. It 
leads us to  the problem of one impurity in the TLL. 

For the weak
backscattering, the Lagrangian of this problem can be written as
\endmulticols
\widetext
\vspace{-7mm}\noindent\underline{\hspace{87mm}}
\begin{equation}
\int d x {\cal L}_t= \int dx  { v_F \over 2}
\{ 
{1 \over v_F^2} \left({{\partial_t \phi_c(t,x) }
     \over {\sqrt{4 \pi}}}
\right)^2 -
\left({{\partial_x \phi_c(t,x) } \over {\sqrt{4 \pi}}} \right)^2 
\}
- \frac{Y T_L u}{ \pi v_F} \cos(2 V t +
\sqrt{2} \phi_c(t,0))
\label{47}
\end{equation}
\noindent\hspace{92mm}\underline{\hspace{87mm}}\vspace{-3mm}
\multicols{2}
\noindent
where we  rescaled $\phi$ back and introduced a new energy cut-off parameter
$Y T_L$ with dimensionless constant $Y$ which will be specified later. 
Parameter $u$ is
related to the weak reflection coefficient as:$u^2=v_F^2 R(\mu/m)$.
For the strong backscattering the tunneling Hamiltonian approach may be applied
\cite{us3}. It was associated \cite{kf} to
the dual representation using the field $\theta$
mutually conjugated to $\phi:\  
[\theta_{\sigma}(x), \phi_{\sigma}(y)]=i 2 \pi sgn(x-y)$. 
The appropriate Lagrangian reads
\endmulticols
\vspace{-6mm}\noindent\underline{\hspace{87mm}}
\begin{equation}
\int d x {\cal L}_t= \int dx  { v_F \over 2}
\{ 
{1 \over v_F^2} \left({{\partial_t \theta_c(t,x) }
     \over {\sqrt{4 \pi}}}
\right)^2 -
\left({{\partial_x \theta_c(t,x) } \over {\sqrt{4 \pi}}} \right)^2 
\}
- \frac{Y T_L u'}{ \pi v_F} \cos(V t +
\theta_c(t,0)/ \sqrt{2}) 
\label{48}
\end{equation}
\noindent\hspace{92mm}\underline{\hspace{87mm}}\vspace{-3mm}
\multicols{2}
\noindent
with $u'^2=v_F^2 (1-R(\mu/m))$ proportional to the free massive
fermion transmittance and the voltage multiplied by $g$ factor \cite{weiss}.
Both these Lagrangian are, indeed, equivalent \cite{schm} if 
interaction dependent relation between $u$ and $u'$ is met 
\cite{fendley,weiss}. 
The above model (\ref{47}) or(\ref{48}) characterizes the  point 
scatterer of 
any backscattering strength at low energy \cite{kf}. Although, the exact relation 
between $u$ or $u'$ and the bare parameters of the scatterer
remains unknown. 
Our problem is dually symmetrical to that of Kane and Fisher:
suppression of the direct current in their problem equals suppression of 
the backscattering one in our case. This correspondence allows us to 
re-write their 
solution\cite{fendley,kf} as follows:
\begin{eqnarray}
I&=&{T_x \over \pi} Im \psi ({1 \over 2}+{T_x + i V\over \pi T})
\nonumber\\
I&=&{T_x \over \pi}\arctan(V/T_x), \ \ T=0 \label{411} \\
G&=&\frac{T_x}{\pi^2 T} \psi ' ({1 \over 2}+{T_x \over \pi T})
\nonumber
\end{eqnarray}
where $\psi$ denotes the digamma function and satisfies: $\psi'(1/2)=\pi^2/2,
\  
\psi'(x) \propto 1/x, \ x \rightarrow \infty$, and
a new energy scale $T_x$ \cite{weiss} 
varies from $T_x=Y T_L/ \sqrt{4R}$ at the weak backscattering (\ref{47})
to  $T_x=Y T_L (1-R)/\pi$ at the strong one (\ref{48}).

Let us, first, compare this result with the
perturbative one \cite{us} of the previous section. The latter was derived making use of 
the long-time asymptotics for the correlator (\ref{6}):
\endmulticols
\vspace{-7mm}\noindent\underline{\hspace{87mm}}
\begin{equation}
<e^{i \phi_c(x,t)} e^{-i \phi_c(y,0)}>=const \bigl( {\alpha
T_L } \bigr)^{2g} \bigl({{(\pi T/T_L )^2} \over {
sinh(\pi (t-i\alpha)T) sinh(\pi (-t+i\alpha)T)}} \bigr)
F(x)F(y)
\label{49} 
\end{equation}
\noindent\hspace{92mm}\underline{\hspace{87mm}}\vspace{-3mm}
\multicols{2}\noindent
where $\alpha=1/E_F$ and $F$ was simplified in the previous section as: 
$F(x)=(x/L)^{gr}\prod_{\pm,m=1}^\infty(m\pm x/L)^{gr^{2m\pm 1}} \approx 
const' \times [x(L-x)/L^2]^{gr}$ and $const'=e^{\gamma(\infty)}$. One can see that substitution of this asymptotics in the 
whole formal series for the backscattering current discussed above implies
transformation of the Lagrangian (\ref{41}) into the one of (\ref{47})
with the coefficient for the $\cos$-term:
$e^\gamma \Gamma(1+r) m J_{r+1/2}(E_{thr}t_L)/\{\sqrt{\pi}
(2E_{thr}t_L)^{r+1/2}\}$ 
instead of $\frac{Y T_L u}{ \pi v_F}$ 
and another energy cut-off $T_L$. This model would be equivalent to that
we constructed before in the weak perturbative regime 
if we can meet 
\begin{equation}
{Y \over \sin(\mu t_L)}=
{2e^{-\gamma} \over \sqrt{\pi} \Gamma(1+r)}
\frac{(2 \mu t_L)^{r-1/2}}
{J_{r+1/2}(\mu t_L)}
\label{410}
\end{equation}
At zero $r$ it exactly specifies $Y$ as a constant on the
order of 1. However, if $r\neq 0$ ($r=1/3$ for $g=1/2$ we assumed), $Y$ increases
$\propto(\mu t_L)^r $ at large $\mu t_L$. Moreover, there is a mismatching 
between the oscillating 
structures of $J_{r+1/2}(\mu t_L)$ and $\sin(\mu t_L)$ 
which cannot be naturally
accounted for by a smooth variation of the energy cut-off, but sooner by a small
deviation of the traversal time $t_L$ in the free electron reflection
coefficient (\ref{44}) from its bare value $L/v_c$ as $\mu$ changes.
Such behavior results from penetration of the interaction inside
the wire. It is described by the finite reflection coefficient $r$ in the 
inhomogeneous TLL model. 
The phenomenon is more important for $\mu \gg max(T_L,m)$ when the electron 
propagation through the wire is not suppressed. In the opposite regime of 
small $|\mu|/T_L<1$ no interference structure is expected and $Y$ remains
constant. Finally, under this choice (\ref{410}) of $Y$  one can see that
$T_x \gg T_L$. Therefore, the 
solution (\ref{411}) coincides
with the perturbative result (\ref{11}, \ref{12})
that is $I-V/\pi \propto -V^3$ and $G -1/\pi \propto -T^2$.
\begin{figure}[htb]
\begin{center}
\leavevmode
\psfig{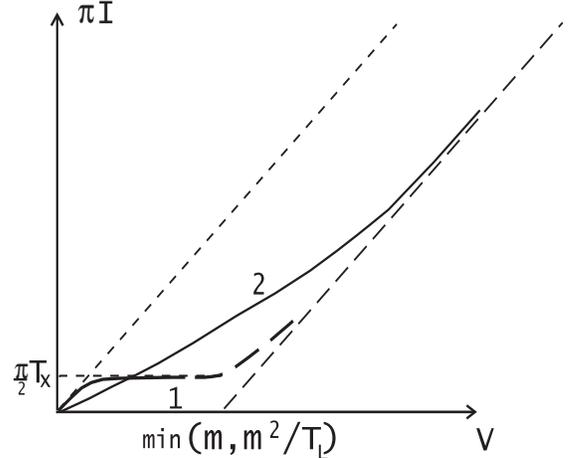}
\narrowtext{
  \caption{Schematic current $I$ vs. voltage $V$ near half-filling
           $\mu \ll T_L$: curve 1 is zero temperature 
           dependence, curve 2 is the high temperature $ T \gg T_L$ one,
           the dashed lines are the low voltage $I=V/\pi$ and 
           high voltage asymptotics.     \label{cn}
        }}
\end{center}
\end{figure}
Turning to the case $m/T_L \geq 1$ we cannot use the perturbative 
expression (\ref{410}) anymore: 
The perturbative series is not convergent due to a finite gap creation.
Then above non-perturbative consideration is necessary.
Application of the solution (\ref{411}) in this case reveals 
a quite remarkable property of low energy transport through 
the Mott-Hubbard insulator. There is
an exponentially small value of 
$T_x=(1-R(\mu/m))YT_L/\pi \propto T_L exp(-2m/T_L)$ 
for $\mu \ll m$. Hence, the zero temperature current $I$ 
(Fig.\ref{cn}) is not 
suppressed for the voltage less than $T_x$ and saturates at $T_x/2$ value 
when $T_x<V<T_L$. Similarly, 
the conductance (Fig.\ref{cty}) displays a small decrease $\propto T^2$ 
below its zero 
temperature value $1/\pi$ with increase of $T$ for $T<T_x$ 
and approaches its exponentially small asymptotics 
$G=T_x/(4T) \propto exp(-2m/T_L)T_L/T$ above $T_x<T<T_L$. As $|\mu |$ increases, 
the reflection coefficient $R(\mu/m)$ on the Fermi level
goes down and $T_x$ exceeds $T_L$, finally, approaching its weak backscattering
value $2T_x=Y T_L/\sqrt{R(\mu/m)}$, where the perturbative consideration
is applicable.

In summary, we studied transport through a 1D Mott-Hubbard insulator
beyond perturbative approach.  Assuming that $g=1/2$ near the half-filling 
in agreement with
the Bethe ansatz solutions we mapped the problem onto 
the exactly solvable models and found current vs. voltage at high
temperature $T>max(m,T_L)$ and at low energy $T,V<T_L$. The solution of
these models shows, in particularly,
that the high-temperature transport through the Mott-Hubbard insulator
is similar to the one through the band gap insulator at $g=1/2$.
At low energies, however, there is always a regime where the transport
remains non-suppressed in the absence of the impurity
backscattering. 
For the strong interaction resulting in the opening
of the large Mott-Hubbard gap,
the transport through the wire is suppressed  near the half-filling almost
everywhere inside the gap except for an exponentially small 
low energy region $V,T < T_L exp(-m/T_L)$.

\section{Acknowledgments}

Most part of this work has been done in collaboration with Naoto Nagaosa when I was exercising
hospitality of the University of Tokyo. I also acknowledge useful discussions  with H. Fukuyama and S. Tarucha.
The work was supported by the Center of Excellence at the Japanese Society for Promotion of Science and by the Swiss National Science Foundation.

\vspace{0.5cm}
$*$ On leave from A.F.Ioffe Physical Technical Institute, 194021, St. Petersburg, Russia

\end{document}